\documentclass[%
 reprint,
 superscriptaddress,
 nobibnotes,
 amsmath,amssymb,
 aps,
prx,
]{revtex4-2}

\usepackage{graphicx}
\usepackage{dcolumn}
\usepackage{bm}

\usepackage{siunitx}
\usepackage{physics}

\begin{document}

\preprint{APS/123-QED}

\title{Machine Learning Inversion from Scattering for Mechanically Driven Polymers }

\author{Lijie Ding}
\affiliation{Neutron Scattering Division, Oak Ridge National Laboratory, Oak Ridge, TN 37831, USA}
\author{Chi-Huan Tung}
\affiliation{Neutron Scattering Division, Oak Ridge National Laboratory, Oak Ridge, TN 37831, USA}
\author{Bobby G. Sumpter}
\affiliation{Center for Nanophase Materials Sciences, Oak Ridge National Laboratory, Oak Ridge, TN 37831, USA}
\author{Wei-Ren Chen}
\email{chenw@ornl.gov}
\affiliation{Neutron Scattering Division, Oak Ridge National Laboratory, Oak Ridge, TN 37831, USA}
\author{Changwoo Do}
\email{doc1@ornl.gov}
\affiliation{Neutron Scattering Division, Oak Ridge National Laboratory, Oak Ridge, TN 37831, USA}

\date{\today}

\begin{abstract}
We develop a Machine Learning Inversion method for analyzing scattering functions of mechanically driven polymers and extracting the corresponding feature parameters, which include energy parameters and conformation variables. The polymer is modeled as a chain of fixed-length bonds constrained by bending energy, and it is subject to external forces such as stretching and shear. We generate a data set consisting of random combinations of energy parameters, including bending modulus, stretching, and shear force, along with Monte Carlo-calculated scattering functions and conformation variables such as end-to-end distance, radius of gyration, and the off-diagonal component of the gyration tensor. The effects of the energy parameters on the polymer are captured by the scattering function, and principal component analysis ensures the feasibility of the Machine Learning inversion. Finally, we train a Gaussian Process Regressor using part of the data set as a training set and validate the trained regressor for inversion using the rest of the data. The regressor successfully extracts the feature parameters.
\end{abstract}

\maketitle


\section{Introduction}
Machine Learning (ML)\cite{murphy2012machine, carleo2019machine} has emerged as a powerful tool for data analysis, enabling the extraction of patterns, trends, and insights from large, complex data sets. Its ability to automate the discovery of meaningful relationships within data has helped to advance numerous fields, including scattering analysis\cite{chang2022machine}. ML techniques can be used for the rapid interpretation of underlying material properties and structural parameters according to complex scattering data. This technique has been applied to various systems including colloids\cite{chang2022machine, huang2023model,tung2023inferring}, copolymers\cite{tung2022small} and lyotropic lamellar systems\cite{tung2024unveiling}.

Polymers are ubiquitous in nature and play a pivotal role in everyday life and for numerous industry settings\cite{de1979scaling, de1990introduction, sperling2015introduction}. Understanding the physics of the polymers can help us to better design and engineer new materials for different applications. The polymers' response to external forces is often of interest as the mechanical properties of the polymer can be revealed accordingly\cite{wang2011salient, li2016stretching,smith1999single,schroeder2005dynamics}. Due to the small physical size of most of the polymers, scattering experiments\cite{murphy2020capillary}, such as X-ray\cite{chu2001small} or neutron\cite{shibayama2011small,chen1986small} scattering, are commonly employed to probe their structure and dynamics at the molecular level. The scattering function measured by these experiments provides indirect but valuable information about the polymer’s conformation and behavior under mechanical stress. Recent advancements in ReoSANS\cite{murphy2020capillary} and sample environments has enabled the application of external forces that compatible to the bending energy of the polymer using flow cells. And a Monte Carlo (MC)\cite{krauth2006statistical} method we recently developed\cite{ding2024off} has enabled the theoretical study of the mechanically driven polymers and calculation of scattering functions comparable to scattering experiments.

Nevertheless, the lack of a scattering analysis technique prevents us from extracting the physical parameters at molecular level from the mechanically driven polymers using small angle scattering experiments. For this we turn to ML for a practical solution. In this paper, we apply a ML inversion technique\cite{chang2022machine} to map the scattering function to inversion targets, or feature parameters of the mechanically driven polymers. We use Gaussian Process Regression (GPR)\cite{williams2006gaussian} to achieve this mapping, and we generate the data set for training and testing using the MC simulation we previously developed. The effects of energy parameters such as bending, stretching, and shear on the scattering function are well reflected independently, and the corresponding polymer deformation is well captured by the calculated scattering function. The feasibility of the proposed ML inversion framework is validated by principal component analysis, which also provide characteristic orientation of the scattering function as a byproduct. Excellent agreement between the ML extracted feature parameters and the MC references are achieved, showing good accuracy for our approach.

\section{Method}
We model the polymer as a chain of $N$ connected bonds with fixed length $l_b$. The tangent of bond $i$ is $\mathbf{t}_i \equiv (\mathbf{r}_{i+1} - \mathbf{r}_i) / l_b$, where $\mathbf{r}_i$ is the position of the joint connecting bonds $i-1$ and $i$. We fix one end of the polymer at the origin. The polymer energy is given by 
\begin{equation}
    E = \sum_{i=0}^{N-2} \frac{\kappa}{2}\frac{(\mathbf{t}_{i+1} - \mathbf{t}_i)^2}{l_b} - \sum_{i=0}^{N-1} (\gamma z_i + f)(l_b \mathbf{t}_i \cdot \mathbf{x})
    \label{equ:energy}
\end{equation}
where $\kappa$ is the bending modulus, $f$ is the stretching force applied in the $x$-direction, $\gamma$ is the shear ratio along the $z$-direction, $z_i = \mathbf{r}_i \cdot \mathbf{z}$ is the $z$-component of the position of joint $i$, and $(\mathbf{t}_i \cdot \mathbf{x})$ is the $x$-component of the bond tangent $\mathbf{t}_i$. A hard sphere interaction between polymer joints, with a sphere radius $l_b/2$, was used to account for self-avoidance of the polymer. 

We sample the configuration space of the polymer using MC and calculate the scattering function and conformation variables of the polymer. We then use Gaussian process regression to achieve a mapping from the scattering function to the system parameters and conformation variables.

\subsection{Monte Carlo simulation}
We simulate the polymer under different system parameters using the Markov Chain Monte Carlo method\cite{ding2024off} we previously developed where each configuration of the polymer is generated by updating the previous one. Two types non-local moves are used for updating the polymer: crankshaft and pivot. Crankshaft executes a random rotation of a inner sub-chain of the polymer while pivot rotates the sub-chain including the end. Details of the simulations can be found in our previous work\cite{ding2024off}. From the simulations, the scattering function and other conformation variables including end-to-end distance, radius of gyration and off-diagonal component of the gyration tensor were computed. The Scattering function is defined as\cite{chen1986small}
\begin{equation}
    I(\vb{Q}) = \frac{1}{N^2} \sum_{i=0}^{N-1}\sum_{j=0}^{N-1} e^{-i \vb{Q} \cdot (\vb{r}_i - \vb{r}_j)},
    \label{equ:2d_scattering_function}
\end{equation}
where $\vb{Q}$ is the scattering vector, and $N$ is the total number of segments. In practice, a projection of the $I(\vb{Q})$ onto a specific plane is collected in the scattering experiments. Since the force field is applied in the $(x,z)$ plane or flow-velocity gradient plane, we calculate the two dimensional $I_{xz}(\vb{Q}) = I(Q_x, Q_y=0, Q_z)$ accordingly. In addition, the end-to-end distance is defined as $R^2 = |\vb{r}_0 - \vb{r}_{N-1}|^2$, the radius of gyration is $R_g^2 = \frac{1}{2}\left<|\vb{r}_i-\vb{r}_j|^2\right>_{i,j}$, and $xz$ the component of gyration tensor is $R_{xz}=\frac{1}{2}\left<(x_i-z_j)^2\right>_{i,j}$, with $\left<\dots\right>_{i,j}$ denoting the average over all pairs of joints.

\subsection{Gaussian process regression}
To obtain a mapping from the scattering function $\vb{x}=I_{xz}(\vb{Q})$ to inversion targets $\vb{y}$ including both system parameters and conformation variables, we train a Gaussian process regressor (GPR)\cite{williams2006gaussian} by feeding training data $\left\{I_{xz}^{train}(\vb{Q})\right\}$ containing scattering functions calculated with various system parameters $(\kappa, f, \gamma)$. Defining the prior on the regression function as a Gaussian process $g(\vb{x}) \sim GP( m(\vb{x}), k(\vb{x},\vb{x}'))$ where $m(\vb{x})$ is the prior mean function and $k(\vb{x},\vb{x}')$ is the covariance function or kernel. Given a test data set $\vb{X}_* = \left\{ I_{xz}^{test}(\vb{Q})\right\}$, the goal of the regressor it to estimate $\vb{Y}_* = g(\vb{X}_*)$. The joint distribution for a Gaussian process is:

\begin{equation}
    \mqty(\vb{Y}~\\ \vb{Y}_*) \sim \mathcal{N}\left( \mqty[m( \vb{X}~) \\ m(\vb{X}_*)], \mqty[k(\vb{X}~,\vb{X}~) & k(\vb{X}~,\vb{X}_*) \\ k(\vb{X}_*,\vb{X}~)~ & k(\vb{X}_*,\vb{X}_*)]  \right)
    \label{equ:Gaussian_process}
\end{equation}
where we use constant prior mean $m(\vb{x})$ and a linear combination of a Radial basis function (Gaussian) kernel and white noise for the kernel $k(\vb{x},\vb{x}') = \exp{\frac{-d(\vb{x},\vb{x}')^2}{2l}} + \sigma \delta(\vb{x},\vb{x}')$, in which $d(\dot, \dot)$ is the Euclidean distance, $l$ is the correlation length, $\sigma$ is the variance of observational noise and $\delta$ is the Kronecker delta function. $l$ and $\sigma$ are the hyperparameters for the regression and can be obtained by training.

\section{Results}
We prepare the training $\left\{I_{train}(\vb{Q})\right\}$ and test sets $\left\{I_{test}(\vb{Q})\right\}$ by carrying out Monte Carlo simulations of the polymer chains with various combinations of energy parameters: bending modulus $\kappa$, stretching force $f$, and shear rate $\gamma$. The scattering function and conformation variables were measured for each simulation. We use natural units in our simulation such that lengths are measured in units of bond length $l_b$ and energies are measured in units of thermal noise $k_B T$. Prior to training, we first study the effect of energy parameters on the scattering function, then validate the feasibility of inversion using principal component analysis. Finally, we train our GPR and compare the ML calculated inversion targets with values calculated using MC.

\subsection{Scattering function of the polymers}
\begin{figure}[!ht]
    \centering
    \includegraphics{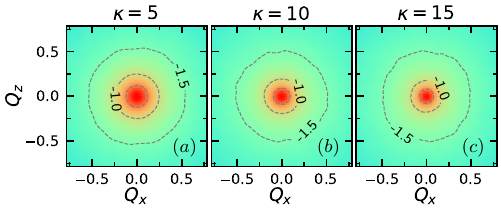}
    \caption{$I_{xz}(\vb{Q})$ of a semiflexible chain with $L = 200$ in its quiescent state with bending modulus $\kappa=5, 10$ and $15$.}
    \label{fig:Iq_kappa}
\end{figure}

\begin{figure}[!ht]
    \centering
    \includegraphics{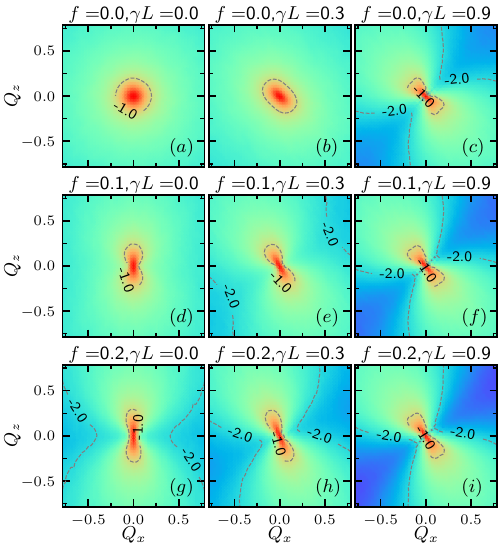}
    \caption{Scattering function $I_{xz}(\vb{Q})$ of a semiflexible chain with $L = 200$ and $\kappa=10$ with various combinations of stretching and shear $(f,\gamma)=(0,0.1,0.2)\times(0,0.3,0.9)$.}
    \label{fig:Iq_fg}
\end{figure}

In order for the GPR to achieve mapping from the scattering function to the inversion targets, the scattering function must reflect the changes of the inversion targets, i.e., the energy parameters. These results are demonstrated in Figs.~\ref{fig:Iq_kappa} and \ref{fig:Iq_fg}, where the scattering function at various bending modulus $\kappa$, stretching force $f$ and shear rate $\gamma$ are shown. 

The bending modulus $\kappa$ determines the persistence length of the polymer. A longer persistence length makes the polymer more rod-like, thus lowers the scattering intensity $I_{xz}(\vb{Q})$ at larger $Q = |\vb{Q}|$. Fig.~\ref{fig:Iq_kappa} shows the $I_{xz}(\vb{Q})$ at different $\kappa$, the contour of $I_{xz}(\vb{Q})$ shows circular symmetry, indicating isotropy of the polymer system in the absence of external forces. The ring of contour level also shrinks as the bending modulus $\kappa$ increases from Fig.~\ref{fig:Iq_kappa}(a) to (c), consistent with our intuition about the effect of the $\kappa$ on the persistence length.

\begin{figure}[!ht]
    \centering
    \includegraphics{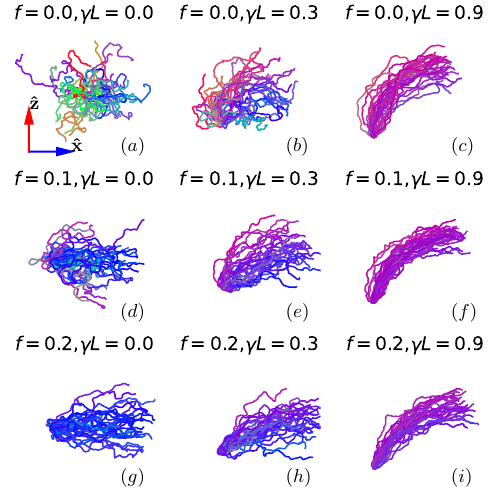}
    \caption{Sample configurations of a semiflexible chain with $L = 200$ and $\kappa=10$ with various combinations of stretching and shear $(f,\gamma)=(0,0.1,0.2)\times(0,0.6,0.9)$, color corresponds to end-to-end orientation in the $xz$ plane. The system is symmetric about $\pm xz$ for (b) and (c) where $f=0, \gamma\neq 0$, these configurations are flipped to the $xz$ direction for better visualization.}
    \label{fig:config_fg}
\end{figure}

When external forces are applied, the polymer deforms accordingly. Fig.~\ref{fig:config_fg} shows sample configurations of the polymer under different stretching $f$ and shear $\gamma$. When only applying stretching $f$ along $x$ direction, the polymer extends along the $x$ as shown in Figs.~\ref{fig:config_fg} (d) and (g). Fig.~\ref{fig:config_fg} (b) and (c) shows that the polymer extends towards the $xz$ direction in a convex manner when only the shear $\gamma$ is applied. Combining the stretching force and shear rate, the polymer behaves like something in the middle, such that an increasing stretching force $f$ pulls the polymer more towards the $x$ direction (compare Figs.~\ref{fig:config_fg}(b), (e) and (f)). These deformations are also reflected in the scattering function. The anisotropic behavior of a polymer should deform the circular symmetric shape of the $I_{xz}(\vb{Q})$. 

Consequently, Fig.~\ref{fig:Iq_fg} shows the corresponding scattering function $I_{xz}(\vb{Q})$ for the polymers corresponding to Fig.~\ref{fig:config_fg}. The contour of scattering function evolves into the oval and then dumbbell shape as applied force increases. The $I_{xz}(\vb{Q})$ at high $Q$ decreases with the increasing magnitude of stretching $f$ and shear $\gamma$, reflecting an increase in the radius of gyration due to straightening. On the other hand, the ratio between $f$ and $\gamma$ affect the orientation of the $I_{xz}(\vb{Q})$ contour. For pure stretching, the contour of $I_{xz}(\vb{Q})$ extend along the $z$ direction, indicating elongation of the polymer along the $x$ direction. In contrast, pure shear makes the contour of $I_{xz}(\vb{Q})$ to extend along the $-xz$ direction, reflecting the elongation of the polymer along the $xz$ direction. Applying and increasing the shear rate on a polymer under stretching, as shown in Fig.~\ref{fig:Iq_fg} (g), (h) and (i), rotates the orientation of the dumbbell shape contour towards the $-xz$ direction.

\subsection{Feasibility of Machine Learning inversion}
Due to the significant difference of the effect on scattering functions induced by different energy parameters, we anticipate that the difference in energy parameters can be distinguished from the scattering function using the GPR. To numerically validate the feasibility of such inversion, we generate $1,680$ random combinations of $(\kappa,f,\gamma)$, in which $\kappa\sim U(2,20)$, $f\sim U(0,0.5)$, $\gamma L\sim U(0,2)$, and $U(a,b)$ is the uniform distribution in interval $[a,b]$, we then run MC simulations to calculate the scattering function $I_{xz}(\vb{Q})$ of the polymer system at these energy parameters, respectively. Each $I_{xz}(\vb{Q})$ is calculated for $2,601 = 51\times 51$ different $(Q_x, Q_z)$ where, $Q_x, Q_z \in [-\frac{50\pi}{L}, \frac{50\pi}{L}]$, uniformly placed on the $51\times 51$ grid. These $I_{xz}(\vb{Q})$ are then flattened to $2601$ dimensional vectors and arranged into a $1,680\times 2,601$ matrix $\vb{F}$. Following a similar ML inversion framework\cite{chang2022machine}, $\vb{F}$ is then decomposed into $\vb{F} = \vb{U}\vb{\Sigma}\vb{V}^T$ using Singular Value Decomposition (SVD)\cite{strang2022introduction}, Such that $\vb{U}$ is $1,680\times 1,680$, $\vb{\Sigma}$ is $1,680\times 2,601$ and $\vb{V}$ is $2,601\times 2,601$. The singular value matrix $\vb{\Sigma}^2$ is diagonal, whose entries are proportional to the variance of the data set $\vb{F}$ projection onto corresponding principal axis\cite{zhu2006automatic}, which is given by the singular vectors $\vb{V}$.

\begin{figure}[!ht]
    \centering
    \includegraphics{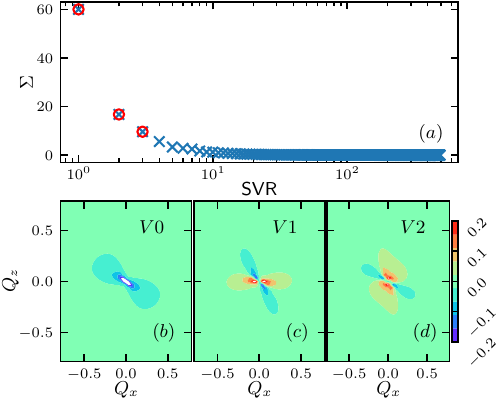}
    \caption{Singular value decomposition (SVD) of scattering function data set. (a) Singular value $\Sigma$ versus Singular Value Rank (SVR), value with top 3 rank are highlighted in red circle. (b)-(d) First 3 singular vectors.}
    \label{fig:SVD}
\end{figure}

Fig.~\ref{fig:SVD}(a) shows the diagonal entries' value of $\vb{\Sigma}$ versus it's Singular Value Rank (SVR). As the SVR increases, it's corresponding value decreases rapidly, indicating the variations in $I_{xz}(\vb{Q})$ are dominated by the first few singular vectors of lower rank. Figs.~\ref{fig:SVD}(b)-(d) shows the first 3 single vectors, which gives a characteristic bases for the $I_{xz}(\vb{Q})$.

\begin{figure}[!ht]
    \centering
    \includegraphics{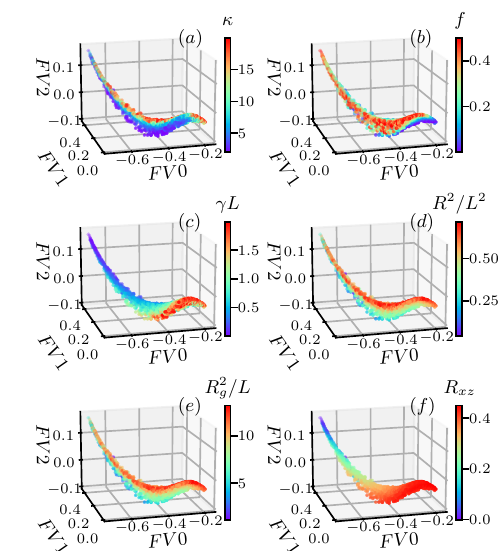}
    \caption{Distribution of various inversion features of training data in the singular value space. (a) Bending modulus $\kappa$. (b) Stretching force $f$. (c) Contour length normalized shear $\gamma L$. (d) End-to-end distance scaled by Contour length square $R^2/L^2$. (e) Radius of Gyration square scaled by Contour length $R_g^2/L$. (f) Off-diagonal, $xz$, component of gyration tensor $R_{xz}$}
    \label{fig:SVD_feature}
\end{figure}

By Projecting the data set $\vb{F}$ onto the first 3 singular vectors $V0$, $V1$ and $V2$. The $(FV0, FV1, FV2)$ coordinates provides a good proxy of the $\vb{F} = \left\{I_{xz}(\vb{Q})\right\}$. Fig.~\ref{fig:SVD_feature} shows the distribution of the 6 inversion targets in the $(FV0, FV1, FV2)$ space. Three of these are the energy parameters: bending modulus $\kappa$, stretching force $f$ and shear rate $\gamma$, another three are conformation variables: end-to-end distance $R^2$, radius of gyration $R_g^2$ and off-diagonal $xz$ component of the gyration tensor $R_{xz}$. In this $(FV0, FV1, FV2)$ space, each point corresponds to one $I_{xz}(\vb{Q})$ in $\vb{F}$, the color represents the corresponding value of inversion targets. From the color distribution, we notice that the inversion targets, feature variables, are all well spread out in the $(FV0, FV1, FV2)$ space, indicating a smooth and continuous mapping between $I_{xz}(\vb{Q})$ and the inversion target can be obtained, thus validating the feasibility of the Machine Learning inversion.

\subsection{Machine Learning inversion of feature variables}
To illustrate the inversion of feature parameters $(\kappa,f,\gamma,R^2,R_g^2,R_{xz})$ from scattering functions $I_{xz}(\vb{Q})$, we partition the total data set $\vb{F} = \left\{I_{xz}(\vb{Q})\right\}$ into two sets, a training set $\left\{I_{xz}^{train}(\vb{Q})\right\}$ consisting $70\%$ of $\vb{F}$, and a test set $\left\{I_{xz}^{test}(\vb{Q})\right\}$ consisting the other $30\%$ of $\vb{F}$. We use the training set to obtain the optimized hyperparameters $(l,\sigma)$, through gradient descent on the log marginal likelihood landscape, for the kernel for each feature parameter individually, and then use the trained GPR with the optimized $(l,\sigma)$ to predict the feature parameters of the test set from their $I_{xz}(\vb{Q})$. The scikit-learn Gaussian Process library\cite{scikit-learn} was used for the training and inversion.

\begin{figure}[!ht]
    \centering
    \includegraphics{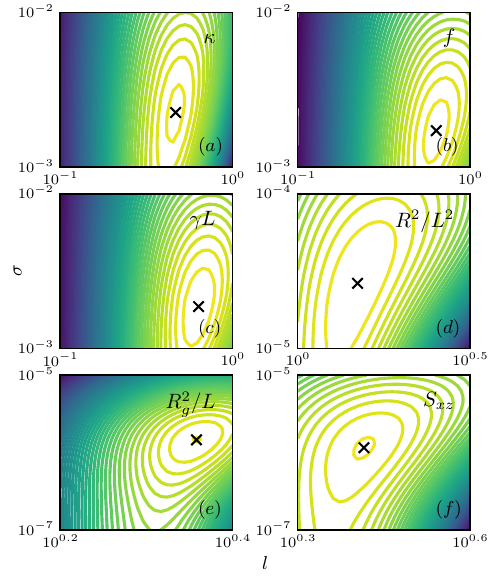}
    \caption{Log marginal likelihood surface of hyperparameters $l$ and $\sigma$ for various inversion features, with optimized value marked with black cross. (a) Bending modulus $\kappa$. (b) Stretching force $f$. (c) Contour length normalized shear $\gamma L$. (d) End-to-end distance scaled by Contour length square $R^2/L^2$. (e) Radius of Gyration square scaled by Contour length $R_g^2/L$. (f) Off-diagonal, $xz$, component of gyration tensor $R_{xz}$}
    \label{fig:LML_contour}
\end{figure}

The log marginal likelihood of the prior is used as the cost function for optimizing the hyperparameters $(l,\sigma)$\cite{williams2006gaussian}. Fig.~\ref{fig:LML_contour} shows the log marginal likelihood contour in $(l,\sigma)$ space for each feature parameter, or inversion targets. The optimized $(l,\sigma)$ are obtained by gradient descent and shown in Tab.~\ref{tab:kernel_parameter}. The contours in Fig.~\ref{fig:LML_contour} show unimodal, convex patterns, which further suggests the reliability of the trained hyperparameters. While the optimized hyperparameters $(l,\sigma)$ differ for each inversion target, two scales of correlation length and noise level emerges. The optimized $l$ and $\sigma$ for all the energy parameters have the same order of magnitude, which is also true for the conformation parameters, but with higher order of magnitude, indicating that the scattering function is more sensitive to the variation of energy parameters comparing to conformation change.

\begin{table}[!ht]
    \centering
    \begin{tabular}{|*{3}{p{0.3\linewidth}|}}
        \hline
          & $l$ & $\sigma$ \\ \hline
        $\kappa$ &  \num{4.6828e-01} & \num{2.2548e-03} \\ \hline  
        $f$ &  \num{6.3714e-01} & \num{1.7219e-03}  \\ \hline                  
        $\gamma L$ & \num{6.3591e-01} & \num{1.8671e-03} \\ \hline 
        $R^2/L^2$ &  \num{1.4921e+00} & \num{2.6388e-05} \\ \hline
        $R_g^2/L$ & \num{2.2814e+00} & \num{1.4582e-06} \\ \hline 
        $R_{xz}$ & \num{2.6027e+00} & \num{1.1527e-06} \\ \hline
    \end{tabular}
    \caption{Optimized hyperparameters for each features, obtained from maximum log marginal likelihood.}
    \label{tab:kernel_parameter}
\end{table}

\begin{figure}[!ht]
    \centering
    \includegraphics{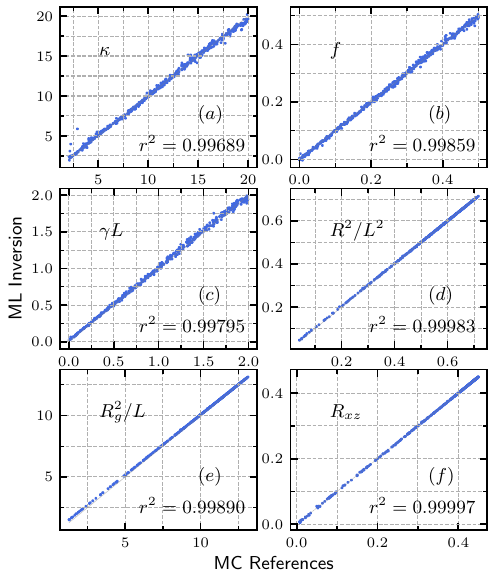}
    \caption{Comparison between the feature variables extracted from the scattering function $I(\vb{Q})$ using the Gaussian Process Regressor and their corresponding computational references calculated using Monte Carlo simulations. Coefficient of determination, $r^2$ scores, are indicated at the bottom of each panel.}
    \label{fig:GPR_prediction}
\end{figure}

Finally, we use the scattering function from test set $\left\{I_{xz}^{test}(\vb{Q})\right\}$ as input to the trained GPR and calculate the feature parameters $(\kappa, f, \gamma, R^2, R_g^2, R_{xz})$ as ML inversion from the $I_{xz}(\vb{Q})$. Fig.~\ref{fig:GPR_prediction} shows the comparison between the GPR predicted feature parameters and the MC references. All of the data lie close to the diagonal line, with $r^2$ score, coefficient of determination, close to 1. The high precision of the inversion shows the power of our Machine Learning approach for extracting important system information from the scattering function.

\section{Summary}
In summary, we apply a ML inversion method to extract feature parameters from the scattering data of mechanically driven polymers. The ML inversion framework was trained based on the theoretically calculated data set of polymer system that is determined by the energy parameters: bending modulus $\kappa$, stretching force $f$ and shear rate $\gamma$. The inversion targets included these energy parameters and conformation variables such as end-to-end distance $R^2$, radius of gyration $R_g^2$ and off-diagonal component of the gyration tensor $R_{xz}$. The scattering function $I_{xz}(\vb{Q})$ of the polymer under different energy parameters was calculated using a MC method we previously developed\cite{ding2024off}. We demonstrate the feasibility of the ML inversion by carrying out PCA of the data set $\vb{F} = \left\{ I_{xz}(\vb{Q})\right\}$ and investigate the distribution of feature parameters by projecting the data set $\vb{F}$ to a 3 dimensional singular vectors space. The GPR was trained and validated, showing that inversion of the feature parameters can be achieved with high-precision.

The versatility of our method promotes its application to the inversion analysis of polymer systems characterized by different intrinsic interactions or under other external forces. For instance, the polymer chains are often charged, in which case instead of using the single parameter, bending modulus, the interaction between monomers on the polymer can be modeled by the two-parameter Yukawa interaction\cite{robbins1988phase}. Moreover, the sample environment of the RheoSANS experiments can introduce nonuniform shear flow like Hagen-Poiseuille flow\cite{batchelor2000introduction}. Furthermore, more complicated polymer systems including polymer brushes\cite{feng2018polymer}, star polymers\cite{ren2016star} and polymer melts\cite{kremer1990dynamics} are also of interest. Modification to the MC simulation can be made accordingly, and ML inversion analysis similar to this work can be carried out.

We note that the inversion method requires the input scattering function to have the same $\vb{Q}$ grid as the training set, which can lead to interpolation of the experimental data in practice. Recent development in ML\cite{liu2024kan} shows possibility of mapping from vectors to functions, which opens a possibility for an alternative way of scattering analysis. Instead of training the mapping from scattering data in discrete $\vb{Q}$ to feature parameters as an inversion, the new framework can learn the mapping from energy parameters to the scattering function in continuous $\vb{Q}$ values, which enables calculation of the scattering function that can then be used for a quick gradient descent optimization of the energy parameter directly. This approach can also be used to cross validate our inversion method.

\begin{acknowledgments}
We thank Jan-Michael Carrillo for fruitful discussions. This research was performed at the Spallation Neutron Source and the Center for Nanophase Materials Sciences, which are DOE Office of Science User Facilities operated by Oak Ridge National Laboratory. This research was sponsored by the Laboratory Directed Research and Development Program of Oak Ridge National Laboratory, managed by UT-Battelle, LLC, for the U. S. Department of Energy. The ML aspects were supported by by the U.S. Department of Energy Office of Science, Office of Basic Energy Sciences Data, Artificial Intelligence and Machine Learning at DOE Scientific User Facilities Program under Award Number 34532. Monte Carlo simulations and computations used resources of the Oak Ridge Leadership Computing Facility, which is supported by the DOE Office of Science under Contract DE-AC05-00OR22725.
\end{acknowledgments}

\bibliography{apssamp}

\end{document}